\documentclass[journal]{IEEEtran}
\ifCLASSINFOpdf
\else
\fi

\usepackage{cite}
\usepackage{subfigure}
\usepackage{amsmath}
\usepackage{autobreak}
\usepackage{amsfonts}
\usepackage{bm}
\usepackage{graphicx}
\usepackage{algorithm}  
\usepackage{algorithmicx}  
\usepackage{algpseudocode}
\usepackage{hyperref}
\usepackage{color}
\usepackage{xcolor}
\usepackage{soul}
\newcommand{\diag}{{\rm diag}}
\newcommand{\Rel}{\mathcal{R}e\Big\{ }
\newcommand{\Rer}{\Big\}}
\newcommand{\sto}{{\rm s.t.}}

\definecolor{bl}{RGB}{0,0,0}
\definecolor{red}{RGB}{0,0,0}
\definecolor{yel}{RGB}{254,249,180}
\newcommand{\bl}[1]{{\color{bl}#1}}
\newcommand{\red}[1]{{\color{red}#1}}

\makeatletter
\renewcommand{\maketag@@@}[1]{\hbox{\m@th\normalsize\normalfont#1}}%
\makeatother

\begin{document}
\title{Optimizing the Age of Information in RIS-aided SWIPT Networks}

\author{Wanting Lyu,~Yue Xiu,~Jun Zhao,~\IEEEmembership{Member,~IEEE},~Zhongpei Zhang,~\IEEEmembership{Member,~IEEE} \\
        
        \thanks{This work was supported in part by the Shenzhen Municipal Natural Science Foundation under JCYJ 20210324140002008 (\textit{Corresponding author:
        		Zhongpei~Zhang}.)
        }
        \thanks{Wanting Lyu and Yue Xiu are with National Key Laboratory of Science and Technology on Communications, University of Electronic Science and Technology of China, Chengdu 611731, China (E-mail: lyuwanting@yeah.net; xiuyue12345678@163.com).

        Jun Zhao is with School of Computer Science and Engineering, Nanyang Technological University, Singapore (E-mail: junzhao@ntu.edu.sg).}
    	
    	\thanks{Zhongpei Zhang is with the Shenzhen institute for Advanced Study, University of Electronic Science and Technology of China, Shenzhen 518110, China(e-mail:zhangzp@uestc.edu.cn).}
 }

\maketitle

\begin{abstract}
In this paper, a reconfigurable intelligent surface (RIS)-assisted simultaneous wireless information and power transfer (SWIPT) network is investigated. To quantify the freshness of the data packets at the information receiver, the age of information (AoI) is considered. To minimize the sum AoI of the information users while ensuring that the power transferred to energy harvesting users is greater than the demanded value, we formulate a scheduling scheme, and a joint transmit beamforming and phase shift optimization at the access point (AP) and RIS, respectively. The \red{alternating} optimization (AO) algorithm is proposed \red{to handle} the coupling between active beamforming and passive RIS phase shifts, and the successive convex approximation (SCA) algorithm is utilized \red{to tackle} the non-convexity of the formulated problems. The improvement in terms of AoI provided by the proposed algorithm \bl{and the trade-off between the age of information and energy harvesting} is quantified by the numerical simulation results.

\end{abstract}

% Note that keywords are not normally used for perreview papers.

% \fi
\IEEEpeerreviewmaketitle

\vspace*{-0.5\baselineskip}
\section{Introduction}
\IEEEPARstart{I}n beyond fifth generation (B5G) and sixth generation (6G) wireless network, challenging requirements are brought about by machine type communications (MTC) and the internet of things (IoT), which include massive connectivity, ultra reliability, low latency, as well as energy efficiency \cite{9210130}. To realize self-sustainable communication systems in future wireless networks, \red{the author in \cite{firstSWIPT} first introduced the concept of simultaneous wireless information and power transfer (SWIPT) theoretically. Recently, SWIPT has becoming popular in the research of wireless communications \cite{SWIPT_review}.}

To achieve the demands for ultra-reliable low latency communication (URLLC) with high quality of service (QoS), \bl{authors in \cite{RealTimeStatus} and \cite{IntroAoI} introduced} a novel metric quantifying the freshness of information, namely the age of information (AoI) for the first times. The AoI is defined as the time elapsed since the generation of the last successfully delivered signal containing status update information about the system.\bl{In addition to traditional communication networks, the AoI has been considered in wireless powered networks \cite{AoI_WPsensor,AgeEH}. In \cite{AoI_WPsensor}, a wireless sensor network was studied with the sensor node which transmits status update packets harvesting energy from radio frequency signals, where the average AoI performance was analyzed and optimized with a greedy method. In \cite{AgeEH}, the authors investigated an age minimization problem in relay-aided two-hop energy harvesting networks. }

\bl{Recently,} reconfigurable intelligent surfaces (RIS) has emerged as a cost-efficient technology to compensate for the propagation attenuation and \red{provide additional links for blocked direct links.} \red{Optimization on RIS aided networks have been investigated widely in the literature.} In \cite{8811733}, the received power \red{was} maximized by jointly optimizing the transmit \red{beamforming} vectors and the phase shifts. Authors \red{in} \cite{9347976} minimized the transmit power \red{under} the constraint \red{of} outage probability requirements. \red{In \cite{8941080}}, RIS \red{was} applied to SWIPT network \red{to maximize} the weighted sum power \red{of} energy harvesting receivers while \red{guaranteeing} signal-to-interference-plus-noise ratio (SINR) \red{of} information users. \red{However, existing works of RIS mainly focused on the improvement of system capacity and energy efficiency. To the best of our knowledge, the investigation on information freshness improvement in RIS-aided SWIPT networks has not been studied yet.}

\red{Driven by this motivation, we study the sum AoI minimization in RIS-aided SWIPT networks.} We construct a framework of RIS-assisted SWIPT network with \red{multiple} information receivers and energy receivers. An non-convex AoI optimization problem is formulated and solved by an successive convex approximation (SCA) based \red{alternating} optimization (AO) algorithm. \red{Numerical results show the significant performance improvement of the proposed algorithm compared with maximum ratio transmission (MRT), random phase shifts case and active amplify-and-forward (AF) relay. Moreover, the trade-off between information freshness and energy harvesting is also highlighted in the simulation results.}

\section{System Model and Problem Formulation}

\subsection{Communication System Model}
\begin{figure}[!t]
    \centering
    \includegraphics[width=0.65\linewidth]{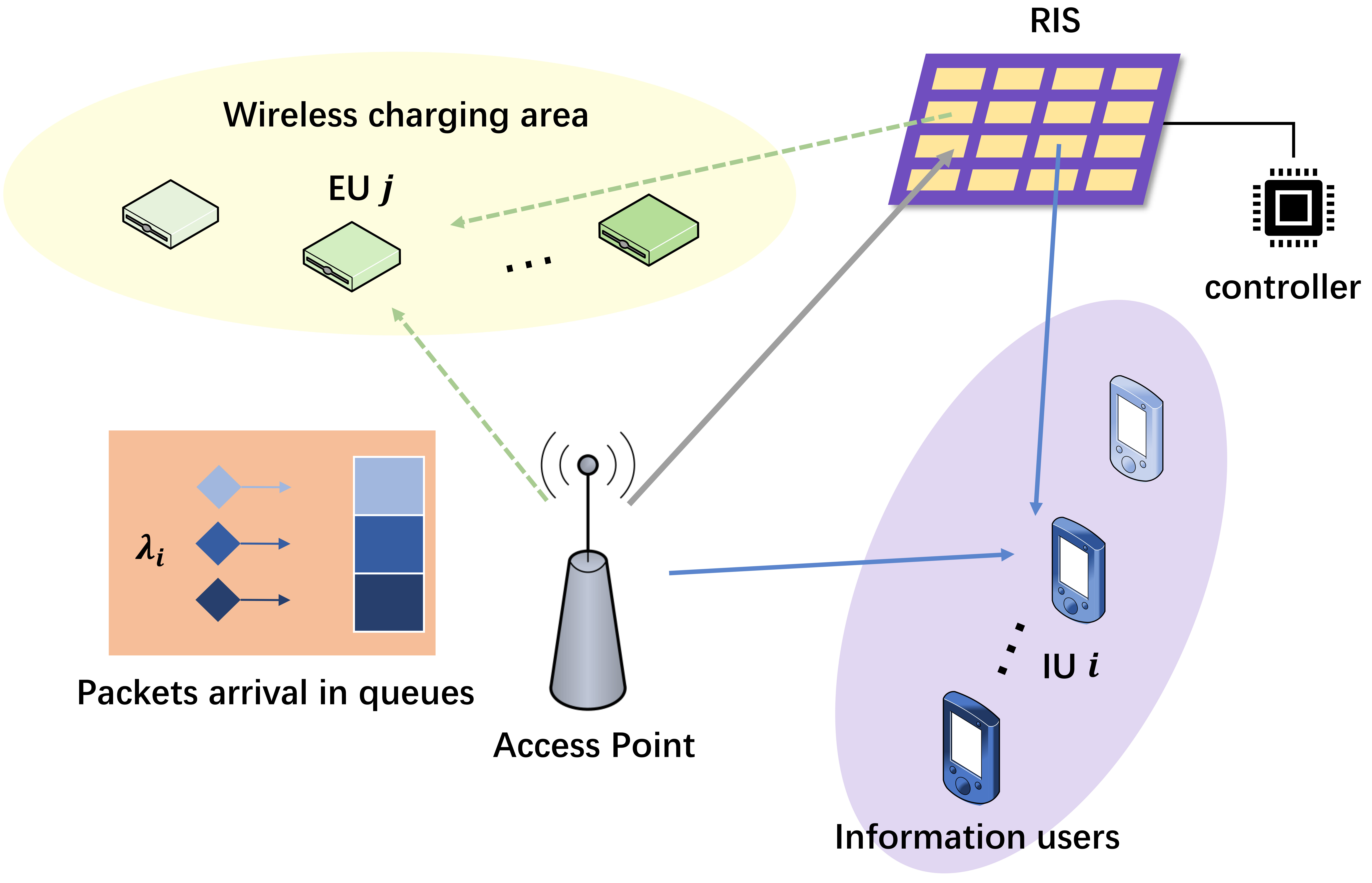}
    \caption{System model for the downlink SWIPT network.}
    \label{model}
\end{figure}

A RIS-aided wireless system is illustrated in Fig. \ref{model}, in which an access point (AP) configured with $N_t$ transmit antennas serves a SWIPT system with the assist of a RIS which has $N_s$ \red{reflecting} elements. We consider $U_I$ information streams at the AP \red{forwarding} status-update signals to the corresponding single-antenna information users (IUs), while $U_E$ energy streams are always forwarding energy packets to the single-antenna energy users (EUs), denoted by the sets $\mathcal{U_I} = \{1,...,U_I\}$ and $\mathcal{U_E} = \{1,...,U_E\}$, respectively. The time \red{axis} is measured by time slots indexed as $t \in \{1,2,...,T\}$. An update packet of information stream $i$ arrives at the system with probability $\lambda_i$ in each slot $t$. The indicator $a_i(t) \in \{0,1\}$ denotes whether the $i^{th}$ information stream has an arrived packet at AP or not.

\red{Assume} a time-varying flat fading channel model with coherent time lasting for one time slot. The channel state information (CSI) \red{in} each time slot is assumed to be \red{perfectly} known. \red{Assume that} the channels for each user are orthogonal without any interference, while the total scheduled information streams cannot exceed the total number of available information channels $M$. Therefore, a scheduling strategy is considered for information streams at the AP, which is constrained by
\vspace*{-0.5\baselineskip}
\begin{gather}
    \sum_i\alpha_i(t) \le M,  \label{chanConst} \\
    \alpha_i(t)\in \{0,1\},  \label{schedIndic}
\end{gather}
\noindent where $\alpha_i(t)$ denotes the scheduling state at time slot $t$ for IU $i$. 

The AP transmit signals separately to IU $i$ and the EU $j$ with $\bm{x}_i^I(t) = \bm{w}_i(t)s_i^I(t)$ and $\bm{x}_j^E(t) = \bm{v}_j(t)s_j^E(t)$, respectively, where $\bm{w}_i \in \mathbb{C}^{N_t\times 1}$ and $\bm{v}_j \in \mathbb{C}^{N_t\times 1}$ are the beamforming vectors for IU $i$ and EU $j$. Information and energy signals are denoted by $s_i^I(t) \sim \mathcal{CN}(0,1)$ and $s_j^E(t) \sim \mathcal{CN}(0,1)$, satisfying $\mathbb{E}(|s_i(t)|^2) = 1$ and $\mathbb{E}(|s_j(t)|^2) = 1$. The total transmit power is limited to be no more than $P_0$, such that
\begin{equation}
    \sum_{i\in \mathcal{U_I}}||\bm{w}_i(t)||^2 + \sum_{j\in \mathcal{U_E}}||\bm{v}_j(t)||^2 \le P_0 \red{.}
    \label{powConst}
\end{equation}

The wireless transmission is assisted by the RIS, where $\bm{h}_{b,i}^H(t), \bm{h}_{r,i}^H(t), \bm{g}_{b,j}^H(t), \bm{g}_{r,j}^H(t) \in \mathbb{C}^{1\times N_t},\text{ and } \bm{G}(t) \in \mathbb{C}^{N_s\times N_t}$ denote the channels between AP and IU $i$, RIS and IU $i$, AP and EU $j$, RIS and EU $j$, AP and RIS, respectively. Define a reflection coefficient matrix $\bm{\Phi}(t) = \diag(e^{j\theta_1(t)}, e^{j\theta_2(t)},..., e^{j\theta_{N_s}(t)}) \in \mathbb{C}^{N_s\times N_s}$, where $\theta_n$ is constrained by 
\vspace*{-0.4\baselineskip}
\begin{equation}
    \theta_n(t) \in [0,2\pi), \forall n \in \{1, 2,..., N_s \},
    \label{phaseConst}
\end{equation}
\noindent which can be equivalently converted to a series of constraints of unit modulus 
\vspace*{-0.4\baselineskip}
\begin{equation}
    |e^{j\theta_n(t)}|=1, \forall n\in \{1, 2,..., N_s \}.
    \label{phaseModu}
\end{equation}

The received SNR for IU $i$ is
$\gamma_i(t) = \frac{|\bm{h}_i^H(t) \bm{w}_i(t)|^2}{\sigma_i^2}$, and the energy harvested by EU $j$ is $E_j(t) = |\bm{g}_j^H(t) \bm{v}_j(t)|^2 $,
\noindent where $\bm{h}_i^H(t) = \bm{h}_{r,i}^H(t)\bm{\Phi}(t)\bm{G}(t) + \bm{h}_{b,i}^H(t)$, and $\bm{g}_j^H(t) = \bm{g}_{r,j}^H(t)\bm{\Phi}(t)\bm{G}(t) + \bm{g}_{b,j}^H(t)$ are the \red{equivalent} channels for IU $i$ and EU $j$.
% At the receiver side, to ensure successive decoding, the receiving SNR is required to satisfy the limitation
% \begin{equation}
%     \gamma_i((\bm{\Phi}(t))) \ge \gamma_{th}, \forall i\in \mathcal{U_I},
% \end{equation}
% while the harvested energy for each EU $j$ should guarantee the constraint

\subsection{The Age of Information}

Following \cite{2021arXiv210306405M}, the AoI for IU $i$ at time slot $t$ is denoted as $A_i(t)$, which is given by
\vspace*{-0.5\baselineskip}
\begin{multline} 
    A_i(t+1) = 
     \alpha_i(t)k_i(t)z_i(t) + (1-\alpha_i(t))(1-k_i(t))A_i(t)+ \\
     (1-\alpha_i(t))k_i(t)A_i(t) + \alpha_i(t)(1-k_i(t))A_i(t) + 1,
     \label{AoIexp}
\end{multline}
% \vspace*{-0.5\baselineskip}
\noindent where $z_i(t)$ is the system time of the packet as follows,
\vspace*{-0.5\baselineskip}
\begin{equation}
    z_i(t+1) = \left \{
    \begin{aligned}
        & 0      & if\, a_i(t+1) = 1, \forall i,t \\
        & z_i(t) + 1 & otherwise.
    \end{aligned}
    \right.
    \label{systime}
\end{equation}
% \vspace*{-0.5\baselineskip}
$k_i(t) \in \{0,1\}$ is $1$ when the scheduled stream $i$ has an available packet, and it turns to $0$ only if the $i^{th}$ stream is scheduled and successfully delivered without any newly arrived packet in this queue. \red{A successful delivery} is considered when the stream is scheduled and the received SNR is no less than the threshold value,
\vspace*{-0.5\baselineskip}
\begin{equation}
    \gamma_i(t) \ge \alpha_i(t)k_i(t)\gamma_{th},\;  \forall i\in \mathcal{U_I}.
    \label{SNRconst}
\end{equation}
\subsection{Problem Formulation}
To minimize the total AoI for $U_I$ information streams, the active beamforming vectors and RIS phase shifts with scheduling schemes are jointly optimized. The optimization problem is proposed as follows,
\begin{align}
    \mathrm{(P1)}: & \min_{\alpha_i(t), \bm{\Phi}(t),\atop \bm{w}_i(t),\bm{v}_j(t)} \sum_t\sum_{i\in \mathcal{U_I}}A_i(t) 
    \label{P1obj} \\
     \sto \quad &(\ref{chanConst}),(\ref{schedIndic}),(\ref{powConst}),(\ref{phaseModu}),(\ref{SNRconst}), \nonumber\\
    & E_j(t) \ge Q, \forall j\in \mathcal{U_E},
    \label{EUconst}
\end{align}
\noindent \red{where (\ref{EUconst}) aims to guarantee the amount of harvested energy.} Due to the fact that the AoI will rise linearly without successful delivery, the problem of minimizing the sum AoI becomes maximizing AoI reduction \red{in each time slot}\cite{9130055}, which is the difference between the AoI and the system time in the current slot. Therefore, the optimization problem is converted to (P2) as
\vspace*{-0.5\baselineskip}
\begin{align}
    \mathrm{(P2)}: & \max_{\alpha_i(t), \bm{\Phi}(t),\atop\bm{w}_i(t),\bm{v}_j(t)} \sum_{i\in \mathcal{U_I}} (A_i(t) - z_i(t))\alpha_i(t)k_i(t)
    \label{P2obj}  \\
     \sto \quad &(\ref{chanConst}),(\ref{schedIndic}),(\ref{powConst}),(\ref{phaseModu}),(\ref{SNRconst}), (\ref{EUconst}). \nonumber
\end{align}

\red{However, (P2) is difficult to solve} due to the non-convexity of the constraints in QoS and \red{harvested} power for the EUs \red{as well as the highly-coupled variables}. To address this issue, \red{an} \red{alternating} optimization (AO) algorithm \red{based on} successive convex approximation (SCA) algorithm is \red{proposed} in the next section.

\section{AO-based Solutions}
In this section, an AO-based algorithm solving this problem is presented by dealing with two subproblems. 
\vspace*{-0.5\baselineskip}
\subsection{Problem Reformulation}
\red{First,} the intractable constraint (\ref{schedIndic}) is relaxed as $\alpha_i(t)\in [0,1]$. \red{Then, to convert the problem into a more friendly form}, define $\rho_n(t) = e^{j\theta_n(t)}$ satisfying $|\rho_n(t)| = 1$, and $\bm{\rho}^H(t) = [\rho_1(t), \rho_2(t),...,\rho_{N_s}(t)]\in \mathbb{C}^{1\times N_s}$. \red{Rewrite} the equivalent channel for IU $i$ as $\bm h_i^H(t) = \bm{\rho}^H(t)\bm{U}_i(t)+\bm{h}_{b,i}^H(t)$, \red{where}
$\bm{U}_i(t) = \diag\Big(\bm{h}_{r,i}^H(t)\Big)\bm{G}(t)$.
Similarly, the equivalent channel for EU $j$ is rewritten as
$\bm{g}_j^H(t) = \bm{\rho}^H(t)\bm{V}_j(t)+\bm{g}_{b,j}^H(t)$. Thus, the problem is rewritten as
\vspace*{-0.5\baselineskip}
\begin{align}
    &\mathrm{(P3)}:  \max_{\alpha_i(t), \bm{\rho}(t),\atop\bm{w}_i(t),\bm{v}_j(t)} \sum_{i\in \mathcal{U_I}} (A_i(t)-z_i(t))\alpha_i(t)k_i(t) 
    \label{P3obj} \\
    &  \sto\quad (\ref{chanConst}),(\ref{powConst}), \nonumber \\
    & \quad \alpha_i(t)\in [0,1],
    \tag{\ref{P3obj}{a}}
    \label{relaschedIndic} \\
    & \quad \big|\rho_n(t)\big| = 1, \forall n\in \{1,2,...,N_s\},
    \label{phaseModu1}
    \tag{\ref{P3obj}{b}} \\
    & \quad \big|(\bm{\rho}^H(t)\bm{U}_i(t)+\bm{h}_{b,i}^H(t))\bm{w}_i(t)\big|^2 \ge \alpha_i(t)k_i(t)\gamma_{th}\sigma_i^2\red{,}
    \tag{\ref{P3obj}{c}}
    \label{SNRconst1} \\
    & \quad \big|(\bm{\rho}^H(t)\bm{V}_j(t)+\bm{g}_{b,j}^H(t))\bm{v}_j(t)\big|^2 \ge Q\red{.}
    \tag{\ref{P3obj}{d}}
    \label{EUconst1}
\end{align}
\vspace*{-1cm}
\subsection{Optimizing the Scheduling Policy and RIS Phase Shifts}

\red{To decompose the highly-coupled problem, we first optimize passive beamforming $\bm{\rho}(t)$ and scheduling indicator $\alpha_i(t)$ with given fixed $\bm{w}_i(t)$ and $\bm{v}_j(t)$}. We employ the penalty method \red{to deal} with the non-convexity of constraint (\ref{phaseModu}) by introducing a large positive constant $C$, where (P3) is reformulated as 
\vspace*{-0.5cm}
\begin{small}
\begin{align}
    \mathrm{(P4)}: & \max_{\alpha_i(t), \bm{\rho}(t)} \sum_{i\in \mathcal{U_I}} (A_i(t)-z_i(t))\alpha_i(t)k_i(t) + C\sum_{n=1}^{N_s}(|\rho_n(t)|^2-1)
    \label{P4obj} \\
    & \sto \quad  (\ref{chanConst}), (\ref{relaschedIndic}), (\ref{SNRconst1}), (\ref{EUconst1}), \nonumber \\
    &\quad |\rho_n(t)| \le 1, \forall n \in \{1,2,...,N_s\}\red{.}
    \tag{\ref{P4obj}{a}}
    \label{phaseModu2}
\end{align}
\end{small}
It is noteworthy that when $C$ starts at a sufficiently small value, a good starting point of $\rho_n(t)$ can be given even though the unit modulus constraints are not \red{satisfied}. Then, we increase $C$ iteratively. With the value of $C$ becoming sufficiently large, the penalty will enforce $|\rho_n(t)| = 1$ to obtain the optimum value (\ref{P4obj}). However, the objective function becomes non-convex \red{due to the penalty term}. To address this issue, SCA algorithm is utilized, where the objective function (\ref{P4obj}) is approximated as the first order Taylor expansion (\cite{distributedRIS}):
\vspace*{-0.8cm}
\begin{small}
\begin{align}
     \sum_{i\in \mathcal{U_I}} &(A_i(t)-z_i(t))\alpha_i(t)k_i(t) + 
     2C\sum_{n=1}^{N_s}\mathcal{R}e\Big\{  \Big(\rho_n^{(i-1)}(t)\Big)^*\Big(\rho_n(t)  \nonumber\\ - &\rho_n^{(i-1)}(t)\Big)\Big\} + C\sum_{n=1}^{N_s}(|\rho_n^{(i-1)}(t)|^2-1),
    \label{Taylobj}
\end{align}
\end{small}
where the value in the $(i-1)^{th}$ iteration is denoted by the superscript $(i-1)$.

Also, non-convex constraints (\ref{SNRconst1}) and (\ref{EUconst1}) can be solved by SCA algorithm, with (\ref{SNRconst1}) being approximated by
\begin{small}
\begin{align}
    & 2\Rel\bm{w}_i^H(t)\Big(\bm{\rho}^{(i-1)H}(t)\bm{U}_i(t)+\bm{h}_{b,i}^H(t)\Big)^H \Big( \bm{\rho}(t) - \bm{\rho}^{(i-1)}(t) \Big)^H \nonumber  \\   &\bm{U}_i(t)\bm{w}_i(t) \Rer 
     + \Big|\Big(\bm{\rho}^{(i-1)H}(t)\bm{U}_i(t)+
    \bm{h}_{b,i}^H(t)\Big)\bm{w}_i(t)\Big|^2 \nonumber  \\
    &\ge \alpha_i(t) k_i(t)\gamma_{th}\sigma_i^2,
    \label{TaySNRalg2}
\end{align}
\end{small}
where the \red{lower bound} of the left hand side (LHS) of the inequality is \red{approximated by} the first Taylor expansion with respect to $\bm{\rho}(t)$. Similarly, (\ref{SNRconst1}) can be estimated as
\vspace*{-0.5\baselineskip}
\begin{small}
\begin{align}
    & 2\Rel\bm{v}_j^H(t)\Big(\bm{\rho}^{(i-1)H}(t)\bm{V}_i(t)+\bm{g}_{b,j}^H(t)\Big)^H \Big( \bm{\rho}(t) - \bm{\rho}^{(i-1)}(t) \Big)^H \nonumber  \\   &\bm{V}_i(t)\bm{v}_j(t)\Rer  
     + \Big|\Big(\bm{\rho}^{(i-1)H}(t)\bm{V}_j(t)+
    \bm{g}_{b,j}^H(t)\Big)\bm{v}_j(t)\Big|^2 \nonumber  \\
    &\ge Q,
    \label{TayEUalg2}
    \vspace*{-0.5cm}
\end{align}
\end{small}
Accordingly, the non-convex problem (P4) is converted as 
\begin{align}
    \mathrm{(P5)}:  &\max_{\alpha_i(t), \bm{\rho}(t)} \sum_{i\in \mathcal{U_I}} (A_i(t)-z_i(t))\alpha_i(t)k_i(t) \nonumber \\
    &+ 2C\sum_{n=1}^{N_s}\mathcal{R}e\Big\{  \Big(\rho_n^{(i-1)}(t)\Big)^*\Big(\rho_n(t) - \rho_n^{(i-1)}(t)\Big)\Big\},
    \label{P5obj} \\
    & \sto \quad  (\ref{chanConst}), (\ref{relaschedIndic}), (\ref{phaseModu2}),(\ref{TaySNRalg2}), (\ref{TayEUalg2}). \nonumber
\end{align}
\noindent where the last term of (\ref{Taylobj}) is omitted since it is a constant.

% \begin{algorithm}[t]  
%   \caption{Scheduling and Phase Shifts Optimization}  
%   \begin{algorithmic}[1]  
%       \State \textbf{Initialize} $ C, \bm{\rho}^{(0)}(t),\alpha^{(0)}_i(t), \bm{w}_i(t), \bm{v}_j(t), \; \forall i \in \mathcal{U_I}$, $\forall j \in \mathcal{U_E}.$ Set iteration index $n = 1$.
%       \Repeat
%       \Repeat
%       \State \textbf{Update} $\alpha_i^{(n)}(t),  \bm{\rho}^{(n)}(t)$  by solving problem (P5).
%       \State Increase $C$.
%       \Until{Unit modulus constraint is satisfied.}
%       \Until{Convergence.}
%       \State \textbf{Output} $\bm{\rho}^{(n)}(t)$.
%   \end{algorithmic}  
% \end{algorithm}
\begin{small}
\begin{algorithm}[t]  
  \caption{Scheduling Scheme and Phase Shifts Optimization}  
  \begin{algorithmic}[1]  
      \State \textbf{Initialize} $ \bm{\rho}^{(0)}(t),\alpha^{(0)}_i(t), \bm{w}_i(t), \bm{v}_j(t), \; \forall i \in \mathcal{U_I}$, $\forall j \in \mathcal{U_E}.$ Set iteration index $n = 1$.
      \Repeat
       \State \textbf{Update} $\alpha_i^{(n)}(t), \bm{\rho}^{(n)}(t)$  by solving problem (P5).
      \Until{Convergence.}
      \State \textbf{Output} $\bm{\rho}^{(n)}(t)$.
  \end{algorithmic}  
\end{algorithm}
\end{small}
Therefore, (P5) becomes a convex problem that can be efficiently solved by standard optimization software such as CVX \cite{boyd2004convex}. The pseudo code \textbf{Algorithm 1} gives a summary of the SCA-based algorithm 1.

\subsection{Optimizing the Scheduling Scheme and Active Beamforming Design}

\red{Given the fixed} phase shift $\bm{\rho}(t)$ obtained by the last sub-problem, \red{we optimize $\bm{w}_i(t)$ and $\bm{v}_j(t)$ while updating $\alpha_i(t)$}. Similarly, SCA algorithm is employed to handle the non-convexity in constraints (\ref{SNRconst1}) and (\ref{EUconst1}), where constraint (\ref{SNRconst1}) is approximated by the first Taylor expansion with respect to $\bm{w}_i(t)$,
\vspace*{-0.2cm}
\begin{small}
\begin{align}
    & 2\Rel\bm{w}_i^{(i-1)H}(t) \bm h_i(t)\bm h_i^H(t) \Big(\bm{w}_i(t) - \bm{w}_i^{(i-1)}(t) \Big) \Rer \nonumber \\   
    &+\big|\bm h_i^H(t)\bm{w}_i^{(i-1)}(t)\big|^2  \ge \alpha_i(t) k_i(t)\gamma_{th}\sigma_i^2,
    \label{TaySNR}
\end{align}
\end{small}
\noindent where $(i-1)$ denotes the value in the $(i-1)^{th}$ iteration. (\ref{EUconst1}) can be rewritten as
\vspace*{-0.2cm}
\begin{small}
\begin{align}
    & 2\Rel\bm{v}_j^{(i-1)H}(t) \bm g_j(t)\bm g_j^H(t) \Big(\bm{v}_j(t) - \bm{v}_j^{(i-1)}(t) \Big) \Rer \nonumber \\   
    &+\big|\bm g_j^H(t)\bm{v}_j^{(i-1)}(t)\big|^2  \ge Q(t).
    \label{TayEU}
\end{align}
\end{small}
\noindent Then, (P3) becomes
\begin{align}
    \mathrm{(P6)}: & \max_{\alpha_i(t),\atop \bm{w}_i(t),\bm{v}_j(t)} \sum_{i\in \mathcal{U_I}} (A_i(t) - z_i(t))\alpha_i(t)k_i(t) 
    \label{P6obj} \\
    & \sto (\ref{chanConst}),(\ref{powConst}), (\ref{relaschedIndic}), (\ref{TaySNR}),(\ref{TayEU}), \nonumber
\end{align}
\noindent which is a convex problem and CVX can be used to solve it \cite{boyd2004convex}. Finally, a stationary point can be obtained by solving a series of approximated convex problems. The corresponding algorithm for scheduling and beamforming optimization is presented in the pseudo code \textbf{Algorithm 2}. As a result, through the \red{alternating} optimization between (P5) and (P6), the value of the objective function will eventually converge to a local optimum.

\begin{algorithm}[t]  
  \caption{Scheduling and Beamforming Optimization}  
  \begin{algorithmic}[1]  
      \State \textbf{Initialize} $ \alpha_i^{(0)}(t), \bm{w}_i^{(0)}(t), \bm{v}_j^{(0)}(t), \bm{\rho}(t), \; \forall i \in \mathcal{U_I}$, $\forall j \in \mathcal{U_E}.$ Set iteration index $n = 1$.
      \Repeat
       \State \textbf{Update} $ \alpha_i^{(n)}(t), \bm{w}_i^{(n)}(t), \bm{v}_j^{(n)}(t) $  by solving problem (P6).
      \Until{Convergence.}
      \State \textbf{Output} $ \alpha_i^{(n)}(t), \bm{w}_i^{(n)}(t), \bm{v}_j^{(n)}(t)$.
  \end{algorithmic}  
\end{algorithm}

\bl{\subsection{Complexity Analysis}
The overall computational complexity is based on the two proposed algorithms. The passive optimization problem is solved by SCA based penalty method, whose complexity is $\mathcal{O}\Big(S_1TU_I^{3.5}\log_2(1/\epsilon_1) \Big)$, where $S_1$ denotes the number of SCA iterations, $T$ denotes the number of iterations to increase $C$, $U_I^{3.5}$ is the complexity of interior-point method in CVX, and $\epsilon_1$ is the accuracy of the SCA algorithm. The complexity of Algorithm 2 is $\mathcal{O}\big(S_2N_s^{3.5}\log_2(1/\epsilon_2) \big)$. Hence, the overall computational complexity of the proposed AO algorithm is $\mathcal{O}\Big(S_A(S_1TU_I^{3.5}\log_2(1/\epsilon_1) + S_2N_s^{3.5}\log_2(1/\epsilon_2) )\Big)$, where $S_A$ is the number of alternating iterations.
}

\section{Numerical Results}

To \red{quantify} the \red{effectiveness of the} proposed algorithm, we perform \red{numerical simulations} with $U_I = 3$ information users and $U_E = 3$ energy users, respectively. We assume $N_t = 4$ transmit antennas at the AP, and single antenna receivers. The packet arriving probabilities $\lambda_i$ are all set as 0.6. The noise power is set to be $-70\ {\rm dBm}$. Besides, $M = 2$ available information channels are supposed.

\begin{figure*}[t]
    \centering
	\subfigure[Sum AoI versus SNR threshold $\gamma_{th}$.]{
		\begin{minipage}[t]{0.29\linewidth}
			\centering
			\includegraphics[width=1\linewidth]{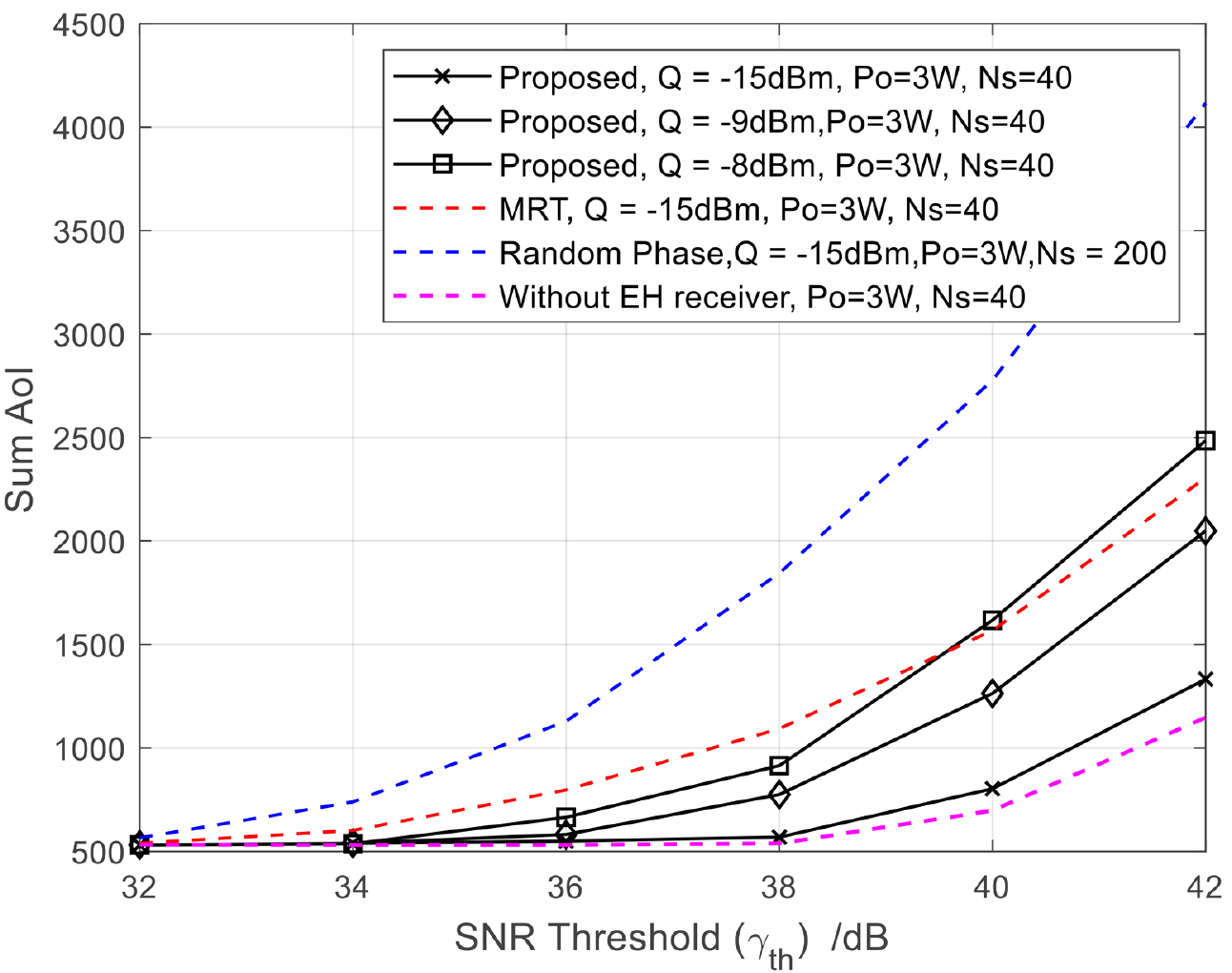}
			\label{AoIvsSNR}
		\end{minipage}
	}
	\subfigure[Sum AoI versus transmit power $P_0$]{
		\begin{minipage}[t]{0.29\linewidth}
			\centering
			\includegraphics[width=1\linewidth]{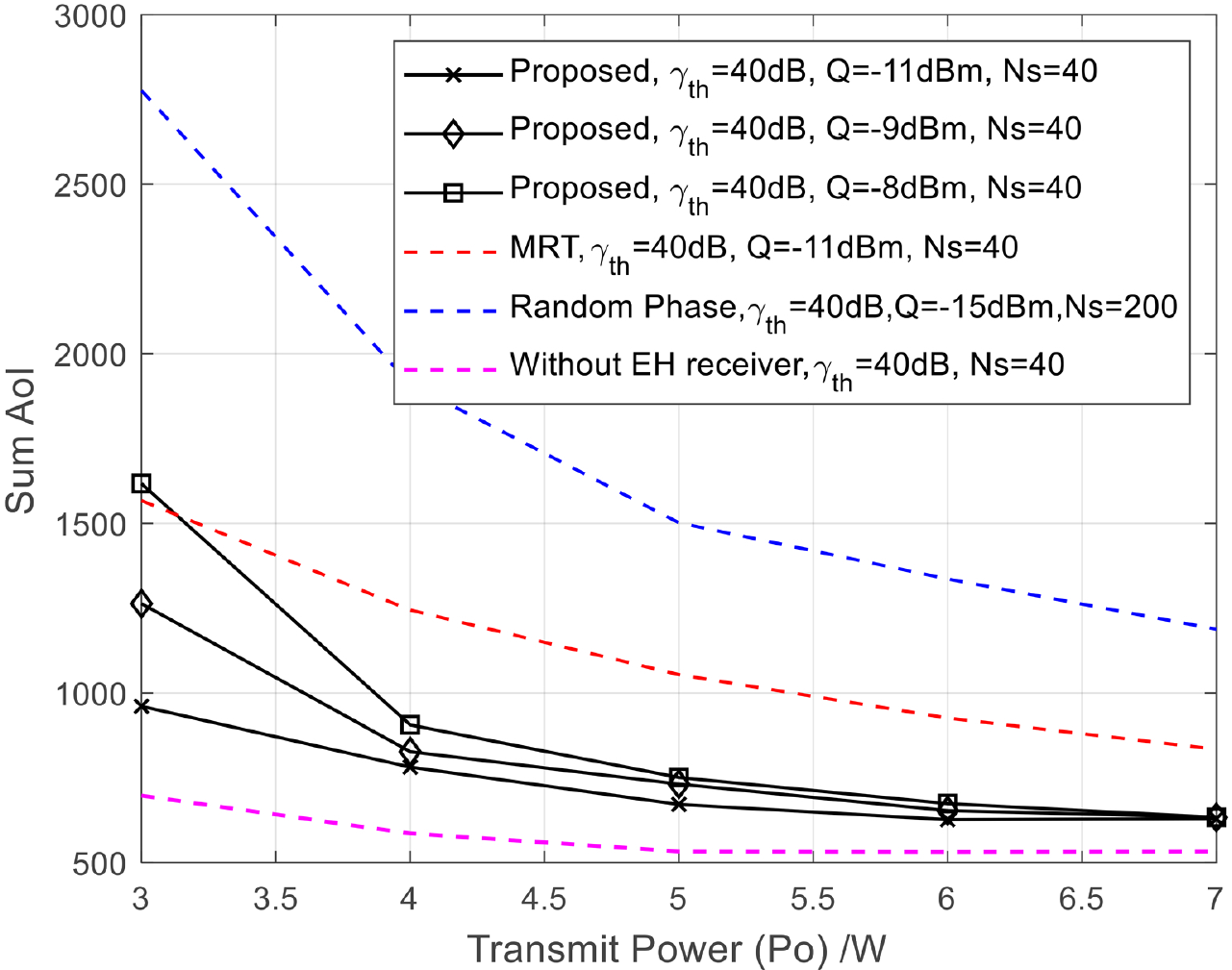}
			\label{AoIvsPo}
		\end{minipage}
	}
	\subfigure[Sum AoI versus the number of reflecting elements $N_s$]{
		\begin{minipage}[t]{0.29\linewidth}
			\centering
			\includegraphics[width=1\linewidth]{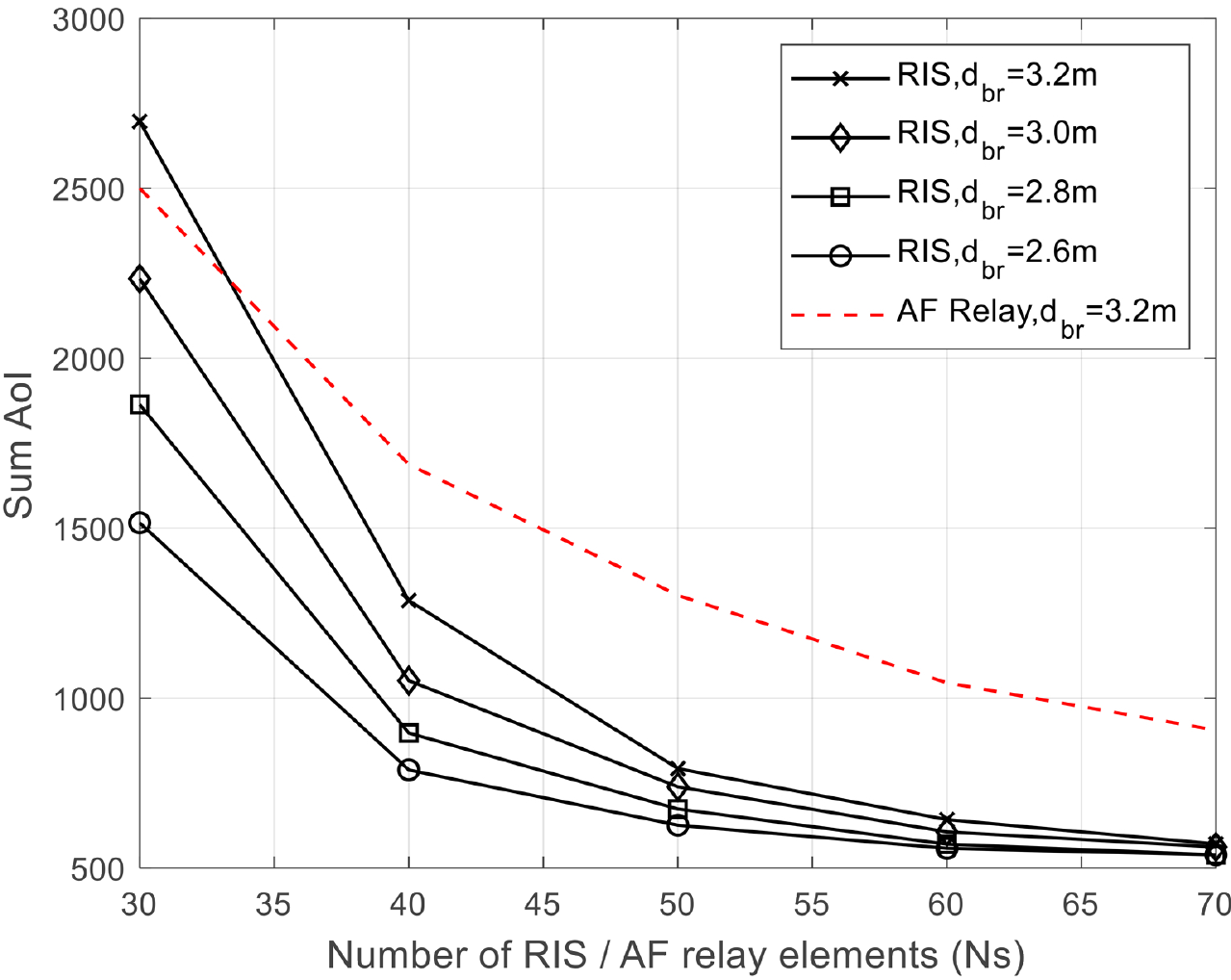}
			\label{AoIvsRIS}
		\end{minipage}
	}
	\caption{Simulation results}
\end{figure*}

We consider Rayleigh fading channels with large scale attenuation, where the independent channels are given as 
\begin{gather}
    \bm{h}_{b,i}(t) = A_{b,i}\bar{\bm{h}}_{b,i}(t) \red{,}\;
    \bm{h}_{r,i}(t) = A_{r,i}\bar{\bm{h}}_{r,i}(t)\red{,}\nonumber\\ 
    \bm{g}_{b,j}(t) = A_{b,j}\bar{\bm{g}}_{b,j}(t)\red{,}\;
    \bm{g}_{r,j}(t) = A_{r,j}\bar{\bm{g}}_{r,j}(t)\red{,}\nonumber\\
    \bm{G}(t) = A_{G}\bar{\bm{G}}(t),
\end{gather}
where $\bar{\bm{h}}_{b,i}(t)$, $\bar{\bm{h}}_{r,i}(t)$, $\bar{\bm{g}}_{b,j}(t)$, $\bar{\bm{g}}_{r,j}(t)$, $\bar{\bm{G}}(t)$ denote the Rayleigh fading channels of the 5 links. $A_{b,i}$, $A_{r,i}$, $A_{b,j}$, $A_{r,j}$, $A_{G}$ denote the corresponding large scale fading, which can be calculated as $A_{b,i} = \sqrt{A_0 d_{b,i}^ {-n_{b,i}} }$ (similar for other channels), where $A_0 = -30  \ {\rm dB}$ is the reference path loss at a distance $d_0 = 1 \text{ m}$, and $n_{b,i} = {n}_{b,j} = 2.2$, $n_{r,i} = n_{r,j} = 2.2$ and $n_{b,r} = 3.5$ are the path loss exponents for AP-IU $i$ link, AP-EU $j$ link, RIS-IU $i$ link, RIS-EU $j$ link and AP -RIS link, respectively. The distances from the AP to the IUs and EUs are set to be $d_{b,i}=31\ \rm m$ and $d_{b,j}=3\ \rm m$, respectively. The AP-RIS distance is set as $d_{br} = 3\ \rm m$ in Fig. \ref{AoIvsSNR} and Fig. \ref{AoIvsPo}.

% **** AoI vs SNR
\red{In Fig. \ref{AoIvsSNR}, the sum AoI increases with rising SNR threshold value $\gamma_{th}$, with the total transmit power set as $P_0 = 3\ \rm W$.} \bl{The 3 solid lines shows the trade-off between the AoI and harvested energy of EUs. The sum AoI increases with growing energy harvesting threshold $Q$ from $-15\ {\rm dBm}$ to $-8\ {\rm dBm}$.} \red{Compared with conventional maximum ratio transmission (MRT) beamforming scheme, the proposed method shows significant AoI performance improvement with the same parameter configuration.} \red{In contrast, sum AoI is larger and  grows faster with increasing SNR threshold using random phase shift even though $Ns$ is set to $200$.} \bl{Moreover, the magenta dash line shows the sum AoI versus $\gamma_{th}$ in the case without EU receiver, from which we can see that sum AoI is lower than SWIPT case, because power allocation and passive beamforming can be adjusted to satisfy QoS of information receivers.
}

% **** AoI vs Po
\bl{Fig. \ref{AoIvsPo} demonstrates sum AoI reduction with growing transmit power. It is because larger transmit power enhances transmit signals, making the SNR and energy harvesting constraints to be more easily satisfied. Furthermore, the AoI and energy harvesting trade-off is also shown in Fig. \ref{AoIvsPo}, where more harvested energy results in worse performance on information freshness. Compared with MRT beamforming scheme and random phase scheme, the proposed AO algorithm shows drastic performance improvement. Compatible with the results in Fig. \ref{AoIvsSNR}, without EH receiver case shows better AoI performance than SWIPT network.
}
% \begin{figure}[!t] % Fig.3
%     \centering
%     \includegraphics[width=0.6\linewidth]{AoIvsPo.pdf}
%     \caption{Sum AoI versus transmit power $P_0$.}
% \end{figure} 

% **** AoI vs Ns
\red{In Fig. \ref{AoIvsRIS}}, we analyze how the number of RIS elements and the distance between the AP and the RIS effect the sum AoI, with $N_s$ increasing from 40 to 70.
The 4 configurations are all set with $P_0 = 3\ \rm W, \gamma_{th} = 40\ {\rm dB} \text{ and } Q = -10\ {\rm dBm}$, while the distance between the AP and the RIS changes between $3.2\ \rm m$ and $2.6\ \rm m$. It can be noticed that with growing number of reflecting elements, the AoI drops quickly to converge, showing the effectiveness of using RIS to enhance the wireless link. Moreover, the shorter distance from the AP to the RIS results in better AoI performance, because the channel gain is larger. In addition, AF relay is used instead of RIS. With the same power consumption, the sum AoI is lower than that of RIS-aided network at the beginning. However, with increasing elements, RIS outperforms AF relay. It is because RIS is passive without radio frequency (RF) chains, and thus RIS-aided networks consumes much lower power than AF relay with RF chains.
% \begin{figure}[!t] % Fig.4
%     \centering
%     \includegraphics[width=0.6\linewidth]{AoIvsRIS.pdf}
%     \caption{Sum AoI versus the number of reflecting elements $N_s$.}
% \end{figure} 

\vspace*{-0.5\baselineskip}
\section{Conclusion}
We studied \red{sum AoI optimization} while satisfying the energy harvesting demands in a SWIPT network with the assistance of RIS. An \red{SCA based} AO algorithm was proposed to \red{solve the scheduling problem with joint active and passive beamforming design. Numerical simulations showed the effectiveness of the proposed method compared with the conventional baselines, and trade-off between AoI and energy harvesting was illustrated by the numerical results.}

\vspace*{-0.5\baselineskip}

\bl{
\appendix[Proof of convergence of the proposed algorithms]

\textbf{Algorithm 1} and \textbf{Algorithm 2} are both based on SCA algorithm, where the convergence proof is similar. Thus, we take \textbf{Algorithm 1} as an example, and convergence of \textbf{Algorithm 2} can be proven as the same manner.

Recall that the problems (P4) and (P5) are equivalent. We show that (P5) can converge to a stationary point, which is equivalent to prove that of (P4). Based on constraints (\ref{SNRconst1}) and (\ref{EUconst1}), we define the following functions 
\begin{small}
\begin{gather}
    \mathcal{F}_1(\bm \rho(t),\alpha_i(t)) = \alpha_i(t)k_i(t)\gamma_{th}\sigma_i^2 - \big|(\bm{\rho}^H(t)\bm{U}_i(t)+\bm{h}_{b,i}^H(t))\bm{w}_i(t)\big|^2,
    \label{Func1} \\
    \mathcal{F}_2(\bm \rho(t)) = Q - \big|(\bm{\rho}^H(t)\bm{V}_j(t)+\bm{g}_{b,j}^H(t))\bm{v}_j(t)\big|^2,
    \label{Func2}
\end{gather}
\end{small}
\noindent where $\mathcal{F}_1(\bm \rho(t),\alpha_i(t)) \le 0$ and $\mathcal{F}_2(\bm \rho(t)) \le 0$ are always guaranteed. With a series of transformations based on SCA algorithm, the above two functions are upper bounded by 
\begin{small}
\begin{align}
    &\mathcal{\Tilde{F}}_1(\bm \rho(t), \bm \rho^{(i-1)}(t),\alpha_i(t)) = \alpha_i(t) k_i(t)\gamma_{th}\sigma_i^2 \nonumber\\ 
    -& 2\Rel\bm{w}_i^H(t)\Big(\bm{\rho}^{(i-1)H}(t)\bm{U}_i(t)+\bm{h}_{b,i}^H(t)\Big)^H \Big( \bm{\rho}(t) - \bm{\rho}^{(i-1)}(t) \Big)^H \nonumber  \\   
    &\bm{U}_i(t)\bm{w}_i(t) \Rer 
     - \Big|\Big(\bm{\rho}^{(i-1)H}(t)\bm{U}_i(t)+
    \bm{h}_{b,i}^H(t)\Big)\bm{w}_i(t)\Big|^2
    \label{Transf_Func1}
\end{align}
\end{small}
\vspace*{-0.5\baselineskip}
\begin{small}
\begin{align}
    &\mathcal{\Tilde{F}}_2(\bm \rho(t), \bm \rho^{(i-1)}(t)) = Q(t)  \nonumber \\
    & -2\Rel\bm{v}_j^H(t)\Big(\bm{\rho}^{(i-1)H}(t)\bm{V}_i(t)+\bm{g}_{b,j}^H(t)\Big)^H \Big( \bm{\rho}(t) - \bm{\rho}^{(i-1)}(t) \Big)^H \nonumber  \\  
    &\bm{V}_i(t)\bm{v}_j(t)\Rer  
     - \Big|\Big(\bm{\rho}^{(i-1)H}(t)\bm{V}_j(t)+
    \bm{g}_{b,j}^H(t)\Big)\bm{v}_j(t)\Big|^2,
    \label{Tranf_Func2}
\end{align}
\end{small}
\noindent which are differentiable. In proposed \textbf{Algorithm 1}, $\mathcal{F}_1(\bm \rho(t),\alpha_i(t))$ and $\mathcal{F}_2(\bm \rho(t))$ are replaced by $\mathcal{\Tilde{F}}_1(\bm \rho(t), \bm \rho^{(i-1)}(t),\alpha_i(t))$ and $\mathcal{\Tilde{F}}_2(\bm \rho(t), \bm \rho^{(i-1)}(t))$, respectively in each iteration of solving (P5). According to \cite{convergence}, the proposed SCA based \textbf{Algorithm 1} converges to a Karush-Kuhn-Tucker (KKT) point of (P4) if the following conditions are satisfied:
\begin{small}
\begin{gather}
\mathcal{F}_1(\bm \rho(t),\alpha_i(t)) \le \mathcal{\Tilde{F}}_1(\bm \rho(t), \bm \rho^{(i-1)}(t),\alpha_i(t))\\
\mathcal{F}_2(\bm \rho(t)) \le \mathcal{\Tilde{F}}_2(\bm \rho(t), \bm \rho^{(i-1)}(t)), \\
\frac{\partial \mathcal{F}_1(\bm \rho(t),\alpha_i(t))}{\partial \bm \rho(t)} = \frac{\partial \mathcal{\Tilde{F}}_1(\bm \rho(t), \bm \rho^{(i-1)}(t),\alpha_i(t))}{\partial \bm \rho(t)}, \\
\frac{\partial \mathcal{F}_2(\bm \rho(t))}{\partial \bm \rho(t)} = \frac{\partial \mathcal{\Tilde{F}}_2(\bm \rho(t), \bm \rho^{(i-1)}(t))}{\partial \bm \rho(t)}.
\end{gather}
\end{small}
Then, the above conditions can be verified by deriving the first-order derivatives of $\mathcal{F}_1(\bm \rho(t),\alpha_i(t))$, $\mathcal{F}_1(\bm \rho(t),\alpha_i(t))$, $\mathcal{F}_2(\bm \rho(t))$ and $\mathcal{\Tilde{F}}_2(\bm \rho(t), \bm \rho^{(i-1)}(t))$ with respect to $\bm \rho(t)$. Notice that (p5) is equivalent to (P4), which means that \textbf{Algorithm 1} can converge to a stationary KKT point of (P4). Moreover, the proof of convergence of \textbf{Algorithm 2} can be similarly derived.

}

\ifCLASSOPTIONcaptionsoff
  \newpage
\fi



\bibliographystyle{IEEEtran}
\bibliography{AoISWIPT}

%\vfill


\end{document}